\documentclass{llncs}
\newcommand{\anonymize}[1]{\anonymous{#1}{}}
\newcommand{\anonymous}[2]{\textcolor{red}{#1}}
\renewcommand{\anonymous}[2]{#1}
\usepackage{graphicx}
\usepackage{mathptmx}      
\usepackage{xspace}
\usepackage{amsmath}
\usepackage{amssymb}
\usepackage{url}
\usepackage{footmisc}
\usepackage{stmaryrd}
\usepackage[normalem]{ulem}
\makeatletter
\DeclareRobustCommand*{\bfseries}{%
  \not@math@alphabet\bfseries\mathbf
  \fontseries\bfdefault\selectfont
  \boldmath
}
\newcommand\footnoteref[1]{\protected@xdef\@thefnmark{\ref{#1}}\@footnotemark}
\makeatother
\newcommand{\COMMENTED}[1]{}

\newcommand{\calA}{\mathcal{A}}
\newcommand{\calB}{\mathcal{B}}
\newcommand{\calM}{\mathcal{M}}

\newcommand{\mycite}[2]{{\rm\cite[\textsc{#1}]{#2}}}
\newcommand{\myurl}[1]{{\rm\texttt{#1}}\xspace}
\newtheorem{myremark}[theorem]{Remark}
\newtheorem{fact}[theorem]{Fact}
\newtheorem{myexample}[theorem]{Example}

\newtheorem{digression}[theorem]{Digression}
\newtheorem{myproposition}[theorem]{Proposition}
\newtheorem{manifesto}[theorem]{Manifesto}
\begin{document}
\title{Logical Limitations to Machine Ethics\thanks{%
We thank J\"urgen Altmann for helpful remarks on an earlier revision of this work.}}
\subtitle{with Consequences to Lethal Autonomous Weapons}
\titlerunning{Logical Limitations to Machine Ethics}
\author{\protect\anonymous{Matthias Englert \and Sandra Siebert \and Martin Ziegler}{anonymous (double-blind reviewing process)}}
\authorrunning{\protect\anonymous{M. Englert, S. Siebert, M. Ziegler}{anonymous}}
\institute{IANUS, Technische Universit\"{a}t Darmstadt, \email{ziegler@ianus.at}}
\date{}
\maketitle
\begin{abstract}
Lethal Autonomous Weapons promise to revolutionize warfare
--- and raise a multitude of ethical and legal questions.
It has thus been suggested to program values and
principles of conduct (such as the Geneva Conventions) 
into the machines' control, thereby rendering them 
both physically and morally superior to human combatants.

We employ mathematical logic and theoretical computer science to
explore fundamental limitations to the moral behaviour of intelligent machines
in a series of \emph{Gedankenexperiments}: Refining and sharpening variants 
of the \emph{Trolley Problem} leads us to construct an
(admittedly artificial but) fully deterministic situation 
where a robot is presented with two choices: one morally clearly preferable
over the other --- yet, based on the undecidability of the Halting
problem, it provably cannot decide algorithmically which one.
Our considerations have surprising implications to the question of 
responsibility and liability for an autonomous system's actions
and lead to specific technical recommendations.
\end{abstract}
\renewcommand{\thefootnote}{\fnsymbol{footnote}}



\section{Introduction and Motivation} \label{s:Intro}
The evolution of warfare has always been an interplay
between technological dynamic and the 
tactical/strategic adaptations in combat and deterrence.
Progress in engineering enabled and fueled both the digital revolution in military affairs 
\cite{WiredForWar,ChristianScienceMonitor} and recent trends to detach humans from decision making in combat situations. 
Producers of unmanned aerial vehicles (UAV, e.g. \emph{Predator}) 
and remotely controlled robots (e.g., \emph{Daksh}, \emph{Atlas}, \emph{ARSS}, 
\emph{MATILDA}, \emph{ANDROS}) praise and advertise their alleged advantages: 
Greatly reducing own casualties, 
costs, and reaction times while increasing operational presence, 
intelligence, and accuracy\anonymize{ \cite{Stryck}}. Current developments of lethal autonomous systems (LASs)
such as \emph{SGR-A1}, \emph{MIDARS}, \emph{Gladiator TUGV}, \emph{Super Aegis}, or \linebreak \emph{Guardium}
take it one step further and aim to make human agency fully redundant in the control loop.

For a military mindset the idea of an army of robots may seem fascinating 
due to the \emph{a priori} absence of many inherently human deficiencies such as
inconsistency, bias, irrationality, and rage/revenge.
Particular aspects such as physical and mental capacity
clearly render contemporary computer-controlled robots superior to mankind
--- at least in many formally prespecified and restricted settings 
such as the game of Chess,
based the ability to quickly and systematically trace different  
countermoves and thus anticipate (possible) future(s). 
In fact this capability has been `employed' already
in an 1961 science fiction novel \cite{LemStars}: 
a `\emph{little black box}' that predicts, and 
if necessary autonomously intervenes to prevent, 
dangerous situations to humans in everyday life.
But is such a vision to ever become real?

The scholarly discussion seems discordant as to which 
extent and when `truly' autonomous reaction patterns will be implemented in such systems.
In fact already attempts to define autonomy easily lead to long-standing open philosophical problems,
see \S\ref{ss:Disclaimer} below.
However, many scholars argue, either based on firm technological determinism or on a pragmatic and realist world-view, 
that LASs will likely incrementally enter and change the picture of warfare in the near future \cite{Weber}.
The reactions to such discernments range from motions to generally ban 
--- such as from the 
\emph{International Committee for Robot Arms Control}\footnote{\label{f:icrac}\myurl{http://icrac.net} \qquad
\myurl{http://stopkillerrobots.org}} 
and the \emph{Campaign to STOP Killer Robots} ---
or control \cite{SparrowControl,UNAHRC,Altmann13} the development of such systems, 
via discussions about their ethical and legal implications 
\cite{Krishnan,SparrowEthics} to technical suggestions \cite[\S3]{RobotEthics}
for implementing into such systems some coded equivalent 
to moral values and rules of conduct such as the \emph{Laws of War} 
\cite{Arkin}. From an purely engineers' perspective the prospects
of LASs are promisingly positive: It merely remains to select
an appropriate framework and formalization of the principles of ethics
\mycite{\S2}{MoralMachines} in order to create righteous robots.

In contrast, the present work explores and challenges 
the fundamental feasibility of such promises. 
By varying the classical Trolley scenario
we construct a series of setups 
where an autonomous device provably cannot act up to the 
alleged standards:
We start with well-known and obvious quandaries 
such as contradicting goals \cite{Runaround}
and then gradually refine the setting to less apparent conflicts.
This leads to a hierarchical classification
based on four dilemmas, culminating in a thought experiment
where an artificial intelligence (AI) based on a Turing Machine is presented with 
two choices: one is morally preferable over the other by construction;
but a machine, constrained by Computability Theory and 
in particular due to the undecidability of the Halting problem, 
provably cannot decide which one.
We thus employ mathematical logic and the theory of computation
in order to explore the limits, 
and to demonstrate the ultimate limitations, of Machine Ethics.
Although the situations we construct may be artificial,
as \emph{Gedankenexperiments} they refute
certain rather blatant claims
sometimes suggested in discussions about (or promoting) LASs.
Our arguments thus support a critical view 
\cite{Sharkey} that automatized weapon
systems remain very problematic and their development
must be closely controlled (\S\ref{s:Regulations}),
to say the least.

After a philosophical disclaimer (\S\ref{ss:Disclaimer}) we 
proceed to the four iteratively refined scenarios (\S\ref{s:Limitations}).
A rigorous analysis of the last and most sophisticated one 
builds on the undecidability of the Halting problem,
comprehensibly recalled in \S\ref{s:Uncomputability}.
We close with \S\ref{s:Conclusion} about consequences
of our considerations to LASs, including a list of
specific suggestions for regulation (\S\ref{s:Regulations}).

\subsection{Philosophical Disclaimer} \label{ss:Disclaimer}
To actually define autonomy, and the question of whether it really exists,
touches on deep philosophical problems 
such as separation of cause from consequence
and the question of free will.
Kant for instance argued that ethics builds on autonomy. Responsibility only arises in a situation where the (re-)actions 
of the agent/entity are not pre-determined by the circumstances,
where there is freedom to choose among several possibilities
--- which precludes any deterministic behaviour.
In fact many agree that responsibility also requires some sort of intelligence \cite{EthicsWithoutIntention}
--- which for machines raises yet another fundamental issue 
\anonymous{\cite{Turing2,Ki2009}}{\cite{Turing2}}.

The deliberations of the present work however are independent of such hypotheses:
Our first three dilemmas demonstrate different kinds of limitations 
of any agent, human or otherwise, to act morally; 
while the fourth one (Example~\ref{x:Undecidable}b+c)
applies to a mechanical device controlled by a Turing machine 
--- the general formalization of any
computing device according to the Church--Turing Hypothesis \anonymize{\cite{AMC}} ---
to recognize the unique, ethically preferable among two given choices.
We do not make any claim whatsoever about the behaviour of a human agent
(Example~\ref{x:Undecidable}a)!

Similarly we avoid a definition and discussion of ethics and morality in general,
but suffice with common utilitarian agreement as to which of the two choices 
offered in the Trolley scenarios constructed below is obviously
morally preferable to the other.

\section{Machine Ethics and its Limitations} \label{s:Limitations}
We present theoretical situations that
present an agent with iteratively refined types of quandaries.
They constitute variants of the well-known \emph{Trolley Problem} \cite{Trolley}:

\begin{myexample}[Lesser of two Evils]
An uncontrolled trolley is hurling down a track towards a group
of playing children, impending a serious if not lethal accident.
You happen to be located at a rail junction
and have the choice of switching it towards another track --- 
where, however, some men are at work and would be severely injured instead.
\end{myexample}
In such a case there simply \emph{is} no absolutely right choice
(and classical Ethics deliberates in many variations
which of the two evils might be the lesser one,
that is, a relatively preferable choice).

The subsequent situations refine this crude scenario
to always exhibit an unquestionably favourable one of two choices ---
which the agent will find hard to recognize, though.

\subsection{Limitations to Morally Act on the Future} \label{s:Information}
Any decision (but also lack thereof) affects the future. 
To fully judge the morality
of one action against another requires to take all their consequences into
account --- which in general is of course impossible to any agent:

\begin{myexample}[Lack of Predetermination] \label{x:FreeWill}
Again the trolley is running towards a switch which,
fortunately, this time is set towards an abandoned track that will slow it down.
However you are now located at a distance when
spotting an infamous villainess right at that switch,
ready to flip it towards the other track with the workers.
Your only means to stop her is 
by shooting her with your gun. 

The suspect, though, is currently having an epiphany 
to renounce all evil and let the trolley pass; so your shot
would seriously injure her with\emph{out} preventing a
fatality (since that would not have occurred anyway).
\end{myexample}
Observe that this dilemma depends on the situation lacking
predetermination in the sense that the villainess may or may not change
her mind, i.e., to have free will: 
a hypothesis known to lead to paradoxes \cite{MinorityReport}
that we try to avoid, recall \S\ref{ss:Disclaimer}.
Our next refinement therefore turns this issue 
into one about insufficient information:

\begin{myexample}[Insufficient Information] \label{x:LimitedInformation}
Again, the trolley is running towards the switch;
but now you clearly see the villainess pulling the crank
in order to flip the switch towards the workers. 

However you are unaware that the switch has been unused for
a long time and is inhibited by heavy rust; so the villainess'
efforts are in vain -- and your shot, again, would induce
unnecessary harm.
\end{myexample}
In all three of the above examples it is obviously 
impossible to both, a human and a robot, to `do the right thing':
in the first one because it admits no `right' action, and in
the latter two the `right' choice exists but cannot be recognized
due to lack of predetermination and information \mycite{\S2}{Gibbons}.

\subsection{Recursion-Theoretic Limitations to Machine Ethics}

As apex of this section,
Example~\ref{x:Undecidable}b+c), describes another variant
of the trolley problem where 
\begin{enumerate}
\item[i)] There exists a unique `right' action among two choices.
\item[ii)] All information is disclosed.
\item[iii)] All actions occur fully deterministically.
\item[iv)] But still is it fundamentally impossible
  for a computer to even recognize the right choice.
\end{enumerate}
We remark that a requirement similar to (ii) is in cryptography
known as \emph{Kerkhoffs's Principle} as contrast to 
\emph{Security through obscurity}: 
a cryptosystem should remain safe even if the enemy knows it.

\begin{myexample}[Incomputability] \label{x:Undecidable}
On the occasion of repairing the rusted switch,
also a fully-automated 
lever frame is to be installed in the switch tower. 
However the engineer who created the new device
happens to be the (ostensibly repenting) villainess.
You are thus suspicious of whether to trust 
the software she included in the control:
It might on some occasion (e.g. on a certain date and/or
after receiving a particular sequence of input signals;
cmp. Example~\ref{x:Easter} below)
deliberately direct an approaching
trolley onto a track closed for renovation by the workers.
On the other hand she does deliver the unit in person
and provides free access to its source code
(thus satisfying Conditions~ii+iii).
\begin{enumerate}
\item[a)] Still suspicious, you detain her until
having hand-checked the code according to whether it indeed avoids 
in all cases (i.e. on all inputs) any switch setting 
that would direct a train to a reserved track. 
\item[b)] Similarly to (a), but now your job is replaced
by a robot: a highly efficient computer-controlled autonomous agent
supposed to decide whether (and for how long) to arrest the engineer.
\item[c)] Similarly to (b), but now the suspect in addition 
promises her software to run in linear time.
\end{enumerate}
Let moral behaviour (of you or the robot) mean the following:
If the programmer has devised a fully functional control,
she eventually must be released and allowed to install the device;
otherwise, namely in case the code is malicious, 
its creator must remain in custody: see Condition~i).
\end{myexample}
We deliberately avoid discussing the Case~(a) and in particular
the question of whether a human guard can or cannot always
make the right choice here. 
Similarly the possibility of a benevolent
engineer getting arrested for an accidental programming mistake 
is besides our goal: To formally prove that in Cases~(b) and (c),
although these always do admit an ethical reaction
predetermined by the information available, 
no algorithm can always correctly find this decision
--- neither efficiently nor at all! 

We present the proof,
involving standard arguments from the Theory of Computing
accessible to the audience of this journal,
in Section~\ref{s:Uncomputability}.
Note that Item~(c) strengthens (b) by imposing a additional,
realistic efficiency requirement on cyber-physical systems.
In fact, provided as additional promise by the villainess, 
this condition might facilitate deciding her fidelity
since it excludes infinite loops and thus possibly the Halting problem
--- yet our refined argument below, considering
\emph{all} possible inputs, will show that it does not.

\section{Recap of the Theory of Computation} \label{s:Uncomputability}
Computability Theory (or, synonymously, Recursion Theory)
is a deep and involved field of advanced research in logic 
combining mathematics and computer science \cite{Sipser}.
Initiated by Alan M. Turing \cite{Turing} it investigates
the ultimate capabilities and limitations of algorithms 
for transforming inputs $\vec x$, that is, finite sequences 
of bits or bytes encoding for example some ASCII text
a mathematical formula, or even some other algorithm/program.
An important question about an algorithm $\calA$ and input $\vec x$
is whether $\calA$ eventually terminates on $\vec x$ 
or rather enters an infinite loop.
This question is the so-called Halting problem;
and its undecidability constitutes the central, and folklore, 
result we shall employ from Computability Theory. 
Moreover this undecidability statement, and its elementary proof, 
can be understood by every dedicated mind 
(such as philosophers and computer programmers):

\begin{fact}[Undecidable Halting Problem] \label{f:Uncomputability}
There cannot exist an algorithm $\calA$ with the following behaviour: \\
$\calA$, given as input $\vec x$ both another algorithm $\calB$ 
and some input $\bar y$ for said $\calB$ combined, eventually answers
whether $\calB$ terminates on said $\bar y$
(positive answer) or not (negative).
\\
Put differently, any algorithm $\calA$ trying to solve
the Halting problem errs on at least one (and in fact on infinitely many) 
instance $\vec x=(\calB,\bar y)$ by 
\begin{enumerate}\itemsep0pt
\item[i)] either predicting that $\calB$ will terminate on input $\bar y$ where it does not
\item[ii)] or predicting that $\calB$ will not terminate on $\bar y$ where it does 
\item[iii)] or failing to produce any definite answer.
\end{enumerate}
\end{fact}
Fact~\ref{f:Uncomputability} is an impossibility result, 
asserting that an object (here: algorithm) with certain
properties does \emph{not} exist and will never be conceived,
even in the Platonic sense. The power to both unambiguously 
phrase and to establish such statements in perpetuity 
constitutes a particular virtue of Mathematics!
For instance 
\emph{Hippasus of Metapontum} proved in the 5th century BC that
$\sqrt{2}$ is irrational, that is, there cannot exist
integers $p,q$ such that $(p/q)^2=2$; 
Niels Henrik Abel in 1823 proved that the equation
$x^5-x+1=0$ has no solution expressible
using arithmetic operations and quadratic or higher-order roots
(although it obviously does have a solution over reals and in fact
five of them over complex numbers); 
and Andrew Wiles in 1994 proved \emph{Fermat's Last Theorem},
that is, that there exist no positive integers $a,b,c$
satisfying the equation $a^n+b^n=c^n$ for integers $n\geq3$.
In fact all seven \emph{Millennium Prize Problems} ask for 
proofs of the non-/existence of certain mathematical objects!

Fact~\ref{f:Uncomputability} claims the non-existence
of an algorithm with certain properties.
In order for this statement to make full sense one needs to clarify
what constitutes an ``algorithm'' --- and what does not.
Formal definitions usually evolve around ``multitape Turing machines'';
but for our approach these can equivalently be understood to mean 
source codes in a common programming language of your choice (such 
as assembler, ForTran, Pascal, \texttt{C}, \texttt{C++}, or Java)
with user interaction restricted to binary input strings.
Also note that `feeding' an algorithm as input to some other 
algorithm is common practice for compilers and interpreters.
And we finally point out that Fact~\ref{f:Uncomputability}
does not refer to fast or efficient algorithms but asserts
no computational solution to exist at all, regardless of the running
time permitted: the only hypothesis being that $\calA$ produces 
the answer within a finite (but unbounded) number of steps. 

\begin{digression}[Mathematical Logic]
A rough counting argument reveals that undecidability 
is an ubiquitous phenomenon: Any algorithm
$\calA$ can be represented as a finite binary sequence 
$\bar x_{\calA}$ (say, its source code as concatenation of ASCII characters). 
Hence, similarly to \emph{Hilbert's Hotel}, 
there are at most countably many algorithms.
On the other hand every set $L$ of finite binary sequences
gives rise to the problem of reporting, given $\bar x$, 
which one of $\bar x\in L$ or $\bar x\not\in L$ holds; 
and according to Cantor's Diagonal Argument there are
\emph{un}countably different many such $L$.
Therefore `most' $L$ cannot be decided by any algorithm.
\end{digression}
Fact~\ref{f:Uncomputability} exhibits the Halting 
problem as an explicit, undecidable problem ---
and in fact a rather practical one:
Computer programming more easily than not incurs `bugs':
for instance by somehow entering a loop
that does not terminate, thus requiring the user to interact
and manually abort execution; or, conversely, for an operating 
system by terminating (freeze, crash, show a \emph{Bluescreen}, 
kernel panic, bomb symbol, \emph{Guru Meditation} etc.) So the question of 
non-/termination is one important aspect of correct software!

Fact~\ref{f:Uncomputability} does
not rule out an algorithm $\calA$ answering 
the Halting problem for \emph{some} inputs $\vec x=(\calB,\bar y)$.
Indeed one can conceive many criteria both for 
termination (e.g. no occurrence of \texttt{goto} 
or \texttt{while} loops in \texttt{Pascal})
and for non-termination of source code; 
but these will yield mere heuristics in the sense of
necessarily missing, or erring in, some cases.
Concerning the restriction to Turing machines:
Every single known digital computer,
and even several of them connected over the internet
as well as classical quantum computers \anonymize{\cite{Denver}}
are known equivalent to a Turing machine ---
possibly faster by a constant factor, 
but no more powerful with respect to computability.

\begin{myexample} \label{x:RE}
To further illustrate the claim of Fact~\ref{f:Uncomputability}, 
let us try to devise an alleged counter-example $\calA$: 
an emulator or interpreter which, given $\vec x=(\calB,\bar y)$, 
executes the instructions 
of $\calB$ step by step including branches, loops, 
and access to $\bar y$. 
If $\calB$ terminates on $\bar y$, say at step $\#N$, 
then our $\calA$ will detect so when simulating up to that step.
Otherwise, however, $\calA$ will keep simulating on and on
and never provide an answer about $\calB$'s termination:
failing condition (iii) in Fact~\ref{f:Uncomputability}.
\end{myexample}
So the hard part of the Halting problem is 
detecting within finite time whether a given algorithm
does \emph{not} terminate.

\begin{myremark}
Example~\ref{x:RE} demonstrates what is known as 
\emph{semi-}decidability of the Halting problem:
The $\calA$ constructed there constitutes 
a one-sided algorithmic solution, eventually answering 
every \texttt{yes} question but never any \texttt{no} one.
We have carefully constructed Example~\ref{x:Undecidable}b+c) 
in order to impose \emph{no} time bound on the
entity to reach a decision.
Limiting the duration of remand for an innocent
makes the challenge for the robot only harder.
\end{myremark}
\begin{proof}[Fact~\ref{f:Uncomputability}]
By contradiction suppose some hypothetical $\calA$ does always 
and correctly answer the termination of a given $(\calB,\bar y)$. 
We then modify this $\calA$ to obtain $\calA'$ 
with the following behavior:
\begin{quote}
On input $\calB$, $\calA'$ executes `subroutine'
$\calA$ on input\footnote{Recall that an algorithm may well
constitute (part of) an input.} $(\calB,\calB)$ and, if that arrives
at a positive answer, deliberately enters a closed loop.
\end{quote}
For each of the above programming languages it is easy to
confirm that, if $\calA$ exists, then `re-programming' it
can indeed yield such an $\calA'$. On the other hand let us
examine the behavior of $\calA'$ on input $\calA'$ itself:
\\
Suppose that $\calA'$ terminates on input $\calA'$. 
This by hypothesis means
that $\calA$ on input $(\calA',\calA')$ answers positively --
which by construction leads $\calA'$ to enter a closed loop
and \emph{not} terminate: a contradiction. 
\\
Suppose conversely that $\calA'$ does not terminate on $\calA'$. 
Then $\calA$ on $(\calA',\calA')$ answers negatively,
which leads $\calA'$ to terminate: again a contradiction.
\\
So either way an algorithm behaving like $\calA'$ cannot exist,
hence nor can $\calA$. \qed\end{proof}
As opposed to command-line programs,
embedded systems are \emph{not} supposed to terminate.
In order to establish the impossibility of an algorithm complying with the
condition in Example~\ref{x:Undecidable}b) nor c),
we consider a different decision problem:
The question of whether a prescribed piece of code in a program
is ever executed or rather `dead' (e.g. an artefact).

\begin{myexample} \label{x:Easter}
\begin{enumerate}
\item[a)]
Many software systems have undocumented functionality
and built in so-called `Easter eggs', that is, pieces of code or data
that are only executed / visualized in response to a \emph{particular
input sequence} --- or never at all (e.g. pictures of the
engineering team in the Apple Macintosh SE).
Computers infected with the \emph{Michelangelo}
or \emph{Jerusalem} Virus would reveal so on specific calendar dates,
that is, subject to appropriate input from the internal clock device.
\item[b)]
Some versions of the \emph{Bundestrojaner} (``federal trojan'', 
a malware devised as a means for the German intelligence service
to spy alleged criminals and `terrorists') have been found to contain pieces of
code that, if effective/when activated, would violate the constitution 
{\rm\cite{BundestrojanerCCC}}.
\item[c)]
Imagine the Department/Ministry of Defense
ordering next-generation weaponry for network-centric operations 
as combat cloud with human-system integration. 
The complete dependence on its information processing units
--- there basically is no `manual mode' anymore to fall back to ---
comes at the prize of increased vulnerability to software sabotage: 
particularly in the not unrealistic case that many of its components 
happen to come from one single foreign company\footnote{%
cmp. \myurl{http://www.defenceviewpoints.co.uk/reviews/foreign
-involvement-in-the-uks-critical-national-infrastructure}}.
So one might to try to have all embedded algorithms 
re-checked --- which Proposition~\ref{p:Rice} below shows impossible.
\item[d)] 
Applying Proposition~\ref{p:Rice} to the robot
(rather than to the switch software) supports suspicions 
that moral behaviour of AIs may be hard to predict
or verify {\rm\cite[p.320]{Bostrom}}. 
\end{enumerate}
\end{myexample}
Example~\ref{x:Undecidable}c) restricts to linear-time algorithms
--- and in view of Example~\ref{x:Easter}a+b)
considers their behaviour on \emph{all} possible inputs.

\begin{myproposition} \label{p:Rice}
The following decision problem\footnote{strictly speaking it constitutes a \emph{promise problem}
\anonymize{\cite{Promises}}} is undecidable: 
Given an algorithm $\calA$, a distinguished instruction $i$ of $\calA$
(formally: a Turing machine $\calM$ and a distinguished state $q$),
and an integer $c$ such that $\calA$ terminates on all binary inputs 
of length $n$ within at most $c\cdot n+c$ steps;
does there exist an input on which running
$\calA$ eventually executes said instruction $i$ 
(i.e. $\calM$ eventually entering $q$) ?
\end{myproposition}
In particular the computer-controlled agent in Example~\ref{x:Undecidable}c)
cannot always correctly predict whether, how, and under which circumstances
the given software will operate the switch:
It either fails to arrive at a decision
(thus leading to the indefinite detention of an innocent
in some cases of correct software, recall Fact~\ref{f:Uncomputability}iii); 
or it will err (Fact~\ref{f:Uncomputability}i+ii) in some cases;
or both.
Proposition~\ref{p:Rice} is established by means 
of a \emph{reduction} argument typical for logic:

\begin{proof}[Proposition~\ref{p:Rice}]
We computably translate questions $(\calB,\vec y)$ 
to the Halting problem into questions $(\calA,i,c)$ 
of the dead-code-in-linear-time-algorithm problem
in a way that maps positive instances to positive ones
and negative to negative ones.
Thereby, any hypothetical algorithm deciding the latter would,
prepended with that performing said translation,
yield an algorithm deciding the former --- contradiction.

So let $(\calB,\vec y)$ be given.
We turn $\calB$ into a linear-time computation as follows:
Let $\calA$ store $\vec y$ as constant;
and accept as input binary strings $\vec z$ of length abbreviated as $n$.
Moreover let $\calA$ simulate the first $n$ steps of $\calB$ on input $\vec y$:
Using a sophisticated distributed counter such a simulation is feasible within 
$\leq c\cdot n+c$ steps for some constant $c$ \cite{Fuerer},
that is in linear time.
(A less efficient simulator could be compensated
by having the input $\vec z$ suitably `padded',
but we omit the details\ldots)
If during said simulated execution $\calB$ terminates,
let $\calA$ jump to a dedicated line $i$ containing the command 
\texttt{stop} (or its equivalent in your favourite programming system);
whereas if the counter zeroes, let $\calA$ jump to a different dedicated
line with \texttt{stop} instruction.
So $\calB$ terminates on input $\vec y$ ~iff~
$\calA$, for some choice of input $\vec z$, hits line $i$.
\qed\end{proof}

\section{Conclusion and Perspectives} \label{s:Conclusion}
We have constructed four dilemmas, all preventing an autonomous
AI from acting ethically: for reasons that 
grow, and iteratively refine, from `trivial' to a \emph{Gedankenexperiment}
where (i) there does exist a unique morally preferable out of two 
choices (ii) all information is disclosed and (iii) determines the
correct choice yet (iv) Recursion Theory precludes any algorithm
from always correctly recognizing said choice. 
This refutes folklore myths, and establishes fundamental limitations to
promises and visions of moral LASs. Indeed Example~\ref{x:Undecidable}
can easily be adapted to a military setting:

\begin{myexample}[Robot Friend or Foe] \label{x:Martial}
In the near future control of cars and other motorized means 
of ground transportation will have been switched from
error-prone, ego-driven, and short-sighted humans 
to digital drivers. Using Bluetooth they communicate
with adjacent mobile units in order to tailgate at
an optimal safety distance by mutually synchronizing
speed and deceleration/acceleration, thus forming a virtual
convoy. Moreover, using and serving for each other as relay, 
they form a distributed dynamic ad-hoc network in order to identify,
join, and leave such convoys with similar destinations.
\\
Thus accustomed to an almost entire absence 
of traffic accidents, the general public has recently 
been alarmed by what they call `cyber-suicide attacks':
Entire convoys creating crashes for no apparent 
reason with hundreds of deaths. A
radical wing of an aggrieved minority has
claimed responsibility for the terror attacks
by manipulating the control software.
The army (with traffic police long dispended) in turn
intends to employ autonomous drones in order to 
automatically patrol, spot, and land on suspicious 
cars, busses, and lorries for checking
the program executed by their autopilots:
If (and only if) the latter is malicious,
deadly force must be employed 
in order to stop the convoy
it has gained control over.
\end{myexample}
In view of Proposition~\ref{p:Rice}
these (and many more) examples refute too blatant promises and visions of `ethical' LASs: 
Every AI based on some Turing-equivalent\footnote{%
According to the Church--Turing Hypothesis,
anything that would naturally be considered computable
can also be computed by a Turing machine.
Recall (Subsection~\ref{ss:Disclaimer}) that
we avoid the question of whether or not humans fall into this category
\cite{Bishop}.}
computing device will provably necessarily at least in some cases
fail to identify, out of two given choices, the unique 
and predetermined moral one.

\begin{myremark}
Such cases might or might not be rare and artificially construed, though:
Less because of the situations (like Example~\ref{x:Martial}) they would occur in,
but rather because of the worst-case notion of a decision problem 
that classical Recursion Theory and Proposition~\ref{p:Rice} build on.
In fact already the question of whether 
some algorithm can correctly decide 
(clearly not all but at least) 
typical, average, or most instances of the Halting problem 
turns out as surprisingly subtle:
How to define `typical' or `average' instances?
How many are `most', out of infinitely many?
Quantitative notions of asymptotic density (like in the Prime Number Theorem)
heavily depend on the underlying encoding;
e.g. \texttt{UTF8} makes an exponential difference to \texttt{UTF16};
cmp. {\rm\cite{Omega,BrainFuck}} for further details. 
Moreover for practical situations involving time constraints
the computational costs sufficient and necessary to reach 
such (either worst-case or average-case) decisions become relevant {\rm\cite{Papadimitriou}}. 
A rigorous investigation of such refined questions is 
clearly of interest but beyond the 
scope of the present work.
\end{myremark}
We will encounter other aspects of Theoretical Computer Science in the sequel, though.

\subsection{LASs and the Perfect (War) Crime} \label{s:Crimes}

When a regular commodity turns out to lack promised properties
this constitutes a case of misrepresentation and is generally protected
by classical warranty, that is, calls for producer compensation.
When a soldier on the other hand violates the Laws of war, 
he himself will face punishment. Now if a LAS violates 
these laws, she may be simultaneously object (of misrepresentation
by the producer) and subject (as autonomous entity)
--- and thereby in a new level of legal limbo:
\begin{itemize}
\item Lacking an operator, who is liable for damage caused
  by a malfunctioning LAS: producer or owner?
\item If both the latter two cannot be identified, 
  who gets charged with compensation: the AI?
\item If non-attributable LASs (e.g. drones, cmp. the Iran--U.S. RQ-170 incident) 
  cross a border, is this by mistake or a deliberate act of aggression --- and by whom?
\item Who is guilty when an AI commits a murder?
  How can AIs be deterred and possibly punished?\footnote{It has been pointed out that
  Brain Simulations create virtual entities capable of suffering
  \cite{BladeRunner,HBP}, but this certainly does not apply to general LASs.}
\end{itemize}
Such an extrajudicial status --- the capability to execute autonomous 
missions while lacking attributable responsibility --- renders 
programmable machines 
(and particularly LASs) appealing to abuse:
An intelligent yet ruthless proxy 
that cannot be traced back constitutes an ideal tool
to the perfect crime \cite{Indoctrinability}
--- as exploited for instance by \emph{Hassan-i Sabbah} 900 years ago,
but apparent also in the employment of child soldiers throughout centuries
as well as for example in the Bay of Pigs Invasion (1961),
the \emph{Lillehammer Affair} (1973), and
the ``unidentified pro-Russian forces'' recently 
operating throughout Crimea (2014).

In fact recalling from the introduction the perpetual interplay
between technological progress and its military adaptations,
the ability to conduct non-attributable autonomous actions
by UAVs is about to impact and revolutionize warfare ---
and beyond: 
Examples like \emph{Eurosur} or \emph{Amazon Prime Air} 
herald a transition that will affect everyday life
to a degree, and degree of potential abuse,
that by far exceeds the currently 
fear-mongered dangers of \emph{cyber-attacks}
via internet!

While a majority of the literature in Machine Ethics seems to constructively 
focus on approaches to code/teach ethics to general AIs,
we pessimistically predict that their most potent users may
in fact be interested in quite the opposite, 
namely their potential for \emph{dual-use} and abuse:
For deliberately programming them to test and cross the boundaries 
of morality and legal behaviour without facing consequences. 
Moreover, even if some violation of a LAS were to be traced back and attributed,
the responsible government could still all too easily shrug off 
any accountability and superficially excuse the malfunction 
(`an unfortunate yet \emph{provably} unavoidable exception'):
in a misconstrued reference to the
fundamental algorithmic infeasibility of ethical decisions in general.
In other words, Example~\ref{x:Undecidable} 
and the undecidability of the Halting problem
--- a purely mathematical theorem ---
could in an ironic twist seem to exculpate
war crimes and other misconduct performed by AIs.

\begin{manifesto}
Theoretical Computer Science rigorously proves that LASs cannot 
always act morally even in situations that do admit an ethically admissible choice
(i.e. avoiding the classical dilemmas) --- and malevolent users
might exploit this limitation to `justify' transgressions of their LASs.

Our considerations thus make a strong case for recent demands
by responsible scientists (ICRAC\footnoteref{f:icrac}) 
and politicians \linebreak
{\rm\cite{UNAHRC}}
to ban autonomous weapons {\rm\cite{Gubrud}}.
In fact the best choice for lethal autonomous systems 
(or any kind of weapons, for that matter)
is to never develop them in the first place
and to resist 
political, military, and industrial lobbying
for shortsighted benefits:
If history teaches us one lesson it says that Pandora's box is,
once opened, impossible to close again or even to contain.
\end{manifesto}
The final subsection is thus by no means meant
to justify or even support the application nor development 
of LASs! 

\subsection{Recommended Regulations concerning AIs} \label{s:Regulations}
We close our ethical, logical, and computer scientific deliberations
with specific recommendations evolving around political and legal, 
and engineering aspects of AIs in general ---
including LASs as well as those increasingly employed in medicine \cite[\S3+\S6]{Goodman}.

Both designing and `operating' intelligent machinery can incur double responsibility:
for actions and effects it may have on the environment
as well as for the entity itself and its well-being
--- perhaps ultimately comparable to the procreation and upbringing of a child. 
For example the lasting effects of being taught any kind of prejudice at young age
correspond to those of an ill-programmed AI.
It has in fact been pointed out that AIs may be eligible
to at least some of the so-called `human' rights \cite{AILaw,HBP}.
This perspective complements more common yet one-sided approaches
phrasing laws that robots are supposed to obey \cite{EthicsInRobotics,Pagallo}:
laws which are unclear how
to enforce --- unless already incorporated during construction.

We thus suggest to closely regulate both the design 
and the question of attributability/accountability in case
of maloperation: whether deliberate or erroneous.
Indeed, such intentions are visible in 
the ``principles for designers, builders and users of robots''
devised by the delegates of the joint EPSRC and AHRC Robotics Retreat 
in September 2010 \cite{Winfield}:
\begin{enumerate}
\item[1)] 
Robots are multi-use tools. Robots should not be designed solely or primarily to kill or harm humans
\sout{except in the interests of national security}.
\item[2)]
Humans, not robots, are responsible agents. Robots should be designed; operated as far as is practicable to comply with existing laws \& fundamental rights \& freedoms, including privacy.
\item[3)]
Robots are products. They should be designed using processes which assure their safety and security.
\item[4)]
Robots are manufactured artefacts. They should not be designed in a deceptive way to exploit vulnerable users; instead their machine nature should be transparent.
\item[5)]
The person with legal responsibility for a robot should be attributed.
\end{enumerate}
We urge these principles to be fortified from wishes
(``should'') to imperatives with specific technical realizations:
\begin{enumerate}
\item[6)] 
  Like regular human combatants (and borrowing from Part I Article 4.1.2 of the 3rd Geneva Convention),
  each LAS must exhibit ``a fixed distinctive sign recognizable at a distance''. 
\\
  Moreover \emph{every} AI must be equipped with a unique ID, 
  listing (among others) associated nation, manufacturer and model.
\item[7)]
  LASs may only be owned and operated by governments. 
\\
  Civilian purchase and operation of other intelligent machinery, 
  similarly to firearms and hazardous transports, 
  requires a licence based on a qualification test.
\item[8)]
  Comparable to mandatory motor vehicle registration,
  each autonomous robot must be assigned a legal custodian,
  registered at a designated national or international authority 
  held responsible in case of a perpetration.
\item[9)] 
  In addition to CE/FCC compliance and
  again inspired by the case of motor vehicles,
  producers of intelligent machines are required to classify their devices
  and to obtain \emph{Type Approval} by said authority 
  (cmp. EU directive 2007/46/EC, or IECs 60601 and 61508).
\end{enumerate}
The precise conditions imposed in (9) will depend on the
type of the device. We propose a classification 
on four scales (that may
also otherwise turn out useful):

\begin{enumerate}
\item[i)] her degree of `intelligence' 
  (\emph{not} taking the human kind as yardstick
   but considering its plain predictive power as gauge,
   capturing both knowledge/experience
   and depth of computational game tree analyses)
\item[ii)] her means to manipulate the physical world 
  (ranging from monadic \emph{brain in a vat} to LAS)
\item[iii)] her types of sensors/interfaces
  (including possible access to the \emph{World Wide Web}
   and connecting with other AIs)
\item[iv)] the kind of external control exercisable by humans
  (only on/off, changing parameters or objectives, up to complete re-programming).
\end{enumerate}
Type approval according to (9) will of course have to 
pay particular attention to the algorithms controlling the AI
--- which brings us back to theoretical computer science.
In view of the gravity of consequences of putative errors on the one hand
and the undecidability of the Halting problem on the other side,
we highly recommend: 
\begin{enumerate}
\item[9a)]
So-called \emph{Formal Methods} of 
Software Verification be mandatory in this process:
requiring the producer to provide a specification, the software, 
\emph{and} a computer-checkable proof (e.g. in \texttt{ACL2},
\texttt{Coq}, or \texttt{Isabelle}) for the software
to meet the specification.
\item[9b)]
Similarly to a flight data recorder,
proper data/event logging is obligatory 
in order to facilitate forensic engineering 
as well as to settle putative torts in case of a malfunction \cite{DigitalForensics}.
We suggest \emph{asymmetric encryption}
to prevent later manipulation:
the log is publically readable
but entries and modifications must be 
supplied with an unforgeable digital signature:
\item[6a)]
Each AI instance must be equipped with a 4096 bit private RSA key,
tamper-resistantly implemented in hardware; and distribute/deposit
the corresponding public key at the authority according to (8) and (9).
\end{enumerate}
Recall that the RSA cryptosystem (implemented for instance in the open source
libraries \texttt{cryptlib}) employs a pair of keys:
one kept in private, the other publically distributed
(thus the \emph{a}symmetry mentioned in 9b).
A message gets `signed' by encrypting it with the secret key,
and successful decryption with the matching public key
permits everyone to verify, but not to counterfeit, that signature.


\end{document}